# Doppler Spectrum Classification with CNNs via Heatmap Location Encoding and a Multi-head Output Layer

Andrew Gilbert, *Student Member, IEEE,* Marit Holden, Line Eikvil, Mariia Rakhmail, Aleksandar Babić, Svein Arne Aase, Eigil Samset, Kristin McLeod

*Abstract*— Spectral Doppler measurements are an important part of the standard echocardiographic examination. These measurements give important insight into myocardial motion and blood flow providing clinicians with parameters for diagnostic decision making. Many of these measurements can currently be performed automatically with high accuracy, increasing the efficiency of the diagnostic pipeline. However, full automation is not yet available because the user must manually select which measurement should be performed on each image. In this work we develop a convolutional neural network (CNN) to automatically classify cardiac Doppler spectra into measurement classes. We show how the multi-modal information in each spectral Doppler recording can be combined using a meta parameter post-processing mapping scheme and heatmaps to encode coordinate locations. Additionally, we experiment with several state-of-the-art network architectures to examine the tradeoff between accuracy and memory usage for resource-constrained environments. Finally, we propose a confidence metric using the values in the last fully connected layer of the network. We analyze example images that fall outside of our proposed classes to show our confidence metric can prevent many misclassifications. Our algorithm achieves 96% accuracy on a test set drawn from a separate clinical site, indicating that the proposed method is suitable for clinical adoption and enabling a fully automatic pipeline from acquisition to Doppler spectrum measurements.

Manuscript received May 31, 2019. This project has received funding from the European Union's Horizon 2020 research and Innovation program under the Marie Sklodowska-Curie grant agreement No 764738.

A. Gilbert and E. Samset are with GE Healthcare and also with the Department of Informatics at the University of Oslo, both Oslo, NO (email: andrew.gilbert@ge.com; eigil.samset@ge.com).

M. Holden and L. Eikvil are with the Norwegian Computing Center, Oslo, NO (email: marit.holden@nr.no; line.eikvil@nr.no).

A. Babić was with GE Healthcare in Oslo, NO for his contribution to this work (email: aleksandar.babic@gmail.com).

M. Rakhmail, S. A. Arne, and K. McLeod are with GE Healthcare in Kharkiv, UA; Trondheim, NO; and Oslo, NO respectively (email: mariya.rakhmayil@ge.com; sveinarne.aase@ge.com; and kristin.mcleod@ge.com).

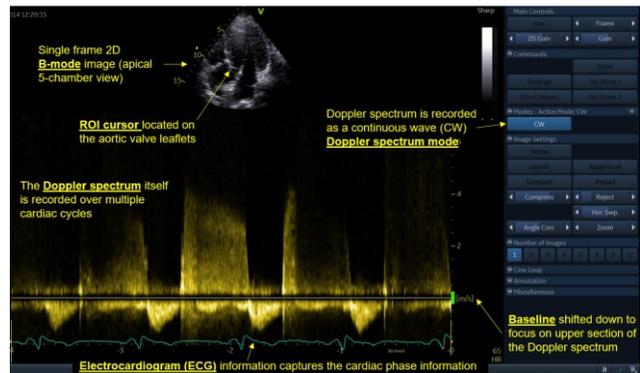

Fig. 1. Example of a doppler acquisition shown on EchoPAC (GE Healthcare, Horten, NO) depicting the relevant information to a spectrum classification problem as a clinician would see it.

## I. INTRODUCTION

ECHOCARDIOGRAPHY is the first point of call when imaging the heart due to its portability, affordability, price, and absence of ionizing radiation. The diagnostic power of echocardiography is reflected in clinical guidelines, with echocardiography indices included as both minor and major clinical diagnostic criteria in many diagnostic protocols [1]. As computational power increases, image quality improves and consequently so does the theoretical accuracy of clinical measurements. In addition to the diagnostic power of echocardiography, there is a growing trend towards further expanding the use of echocardiography as a therapy guidance tool to support interventions and complement other imaging modalities. Minimally invasive valve interventions are becoming the therapy of choice as techniques and prosthetics are advancing, far outweighing the side-effects and risks of full surgery. Techniques to assess blood flow across valves are crucial in therapy planning and follow-up [2]. Spectral Doppler imaging has become an integral component of the echocardiography exam to provide a means to assess hemodynamic function in all four valves of the heart, and therefore has great potential not only for diagnosis but also for therapy planning and interventions.

### A. Spectral Doppler Measurements

Fig. 1 shows an example of a spectral Doppler acquisition as seen in EchoPAC (GE Healthcare, Horten, NO). There are

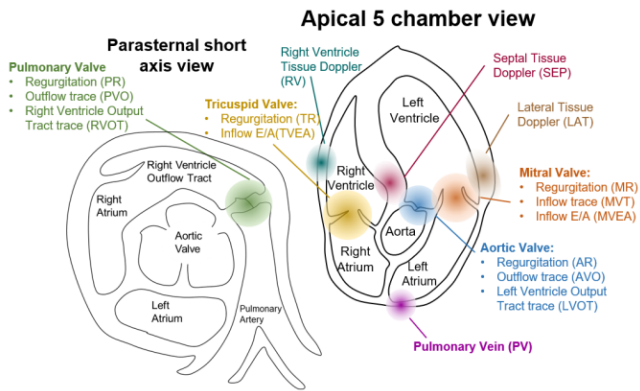

Fig. 2. The location of the ROI for each of the Doppler classes. Although many other views can be (and are typically) used when acquiring the spectra, these two demonstrate the relative location of the classes.

many important features of the acquisition that are available within each recording:

- The **Doppler spectrum** is displayed over multiple cardiac cycles for analysis and measurement.

- The **ECG signal** provides cycle information for orienting temporally with respect to the heart cycle.

- The **relative baseline** of the Doppler spectrum can be adjusted by the user during acquisition to focus on a specific part of the spectrum and prevent aliasing.

- The **mode** provides information on how the Doppler spectrum was acquired. Spectral Doppler incorporates three main imaging modes; continuous wave (CW) Doppler, pulsed wave (PW) Doppler, and tissue Doppler (TVD). CW is used to measure high velocity blood flow across valves, PW provides flow analysis at specific spatial points, and TVD provides quantifiable myocardial velocities.

- The 2D **B-mode** (brightness mode) image shows the orientation of the probe with respect to the physical anatomy of the heart. Doppler spectra can be obtained from a variety of probe positions and angles depending on the desired measurement. The scan converted B-mode image is displayed here to orient the user.

- The **region of interest** (ROI) cursor visible on top of the B-mode indicates where in anatomical space the Doppler spectrum was extracted from. This parameter is interpreted in the context of the B-mode image. See Fig. 2 for a visual overview of how the ROI location corresponds to specific points of a B-mode image, in two example B-mode cardiac views. In the TVD classes the ROI is focused directly on the tissue, while in the CW and PW classes the ROI is focused on an area of blood flow. Exact positioning will depend on the desired measurement, operator preference, and individual patient anatomy.

Together, these pieces of information identify the type of Doppler spectrum.

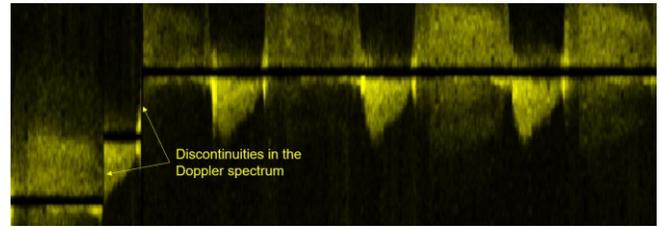

Fig. 3. Discontinuities arise when the operator shifts the baseline during acquisition. This is common practice when acquiring several measurements.

### B. Clinical Need for Automatic Doppler Spectrum Classification

Accurate automatic classification of Doppler spectra can be combined with already available automated measurement techniques (e.g. [3], [4]) to provide fully automated analysis of Doppler spectra. This can increase the efficiency of the clinical workflow, allowing clinicians to spend more time on more difficult measurements.

Furthermore, many clinics have petabytes of patient data in their archive systems from tracking patients over time. Thus, if used in combination with automated measurement techniques, one application of automatic Doppler spectrum classification is to perform rapid historical analysis on past exams in a robust and standardized manner. Since knowing the patient status in reference to previous states can provide further information to support therapy planning, this historical analysis would provide clinical value through objective analysis of measurements over time. An additional application is consistently and continually performing analysis and measurements on groups of patients which could be used, for example, to bring statistical power to the development and augmentation of clinical guidelines.

### C. Previous Work

This problem is unique because of the heterogeneity of data available in each classification example. As shown above, each recording contains image data, spectral data, modal parameters, a baseline position, and ROI coordinate locations. Previously, many of these items have been automatically classified individually, borrowing techniques from non-medical domains. Processing of spectral data has been a common task for several decades in speech recognition [5] and these techniques have been applied to Doppler spectra as well. For example, Wright *et al.* who showed the unique signatures of Doppler spectra from different arteries could be classified with high accuracy by artificial neural networks [6]. Meanwhile, automatic image classification has also become increasingly common as CNNs have achieved super-human performance on many tasks. Recently, these techniques were applied to echocardiographic B-mode images to automatically classify cardiac views with very good results [7], [8].

In non-medical fields, several groups have also looked at how data from different modalities can be combined. Ngiam *et al.* showed how a deep autoencoder could be trained with both video and speech data to generate a shared representation [9]. Ephrat *et al.* demonstrated how video and speech data could be encoded separately and then combined in a bidirectional

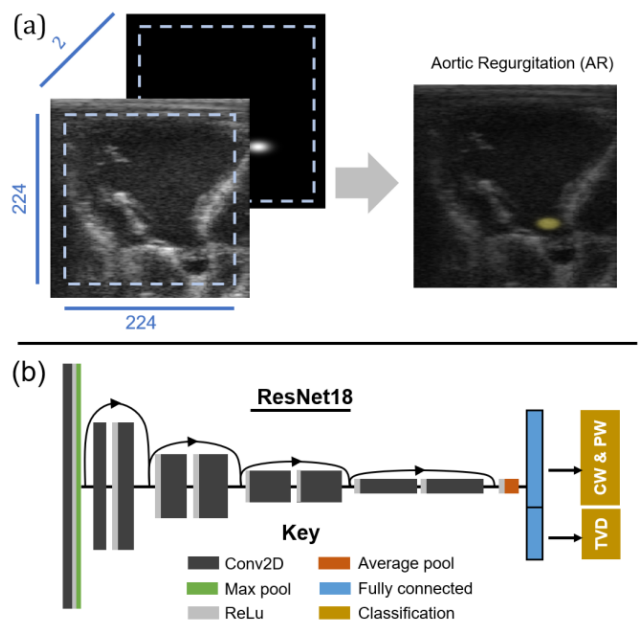

Fig. 4. (a) Heatmap input for our spectrum classification. The heatmap of the ROI location is appended as a channel to the non-scanconverted B-mode image and both are cropped to 224x224 before being input to the network (see section III A). The heatmap is shown overlaid on the B-mode image at right to show the location of the ROI for an example Aortic Regurgitation (AR) case. (b) Network architecture for the given problem following *[23]*. The last fully connected layer is split into two groups based on the mode of the respective class of each node. Each group is passed to its own softmax classifier.

LSTM to solve the cocktail party problem of singling out a single speaker in a noisy audio track [10]. While many deep learning techniques have successfully made the transition from non-medical to medical applications [11], applying multimodal learning techniques remains a challenge because there are several orders of magnitude difference in the amount of data available. Ephrat *et al.* were able to use >2000 hours of automatically annotated data. The annotation of such a volume of data in the context of Doppler spectra is challenging due to the lack of available simulated data. Transfer learning and fine tuning have previously been applied to solve data magnitude problems in medical imaging [12]. However, it is of limited use here since task objectives are different, and in our case the relationship between the modalities (Doppler spectrum to B-mode) varies for each Doppler measurement class.

A challenge in ultrasound imaging, and perhaps medical imaging in general, is images that are acquired in clinical settings are not necessarily standard views. This is particularly a concern in the given spectrum classification problem where misclassifications can be costly, and the model is exposed to only a subset of possible views that might be seen in a clinical workflow. Therefore, an algorithm to classify such images needs a mechanism to handle non-standard cases. This can be either collecting large data-sets that can cover all possible views (even those that are non-standard) or a mechanism to bail-out when the image doesn't fall in the label set, e.g. via confidence metrics with a set threshold for acceptance.

Several groups have looked at how networks can give a prediction of confidence along with an output label. It is well known that CNNs are prone to overfitting and cannot generalize well from the training set to unseen inputs [13]. Previously, Bayesian models have been used to provide a better estimate of model uncertainty by encoding model weights as a probability distribution, but often come with increased parameter count and a higher computational cost to adequately model random distributions [14]. Monte Carlo dropout (MC-dropout) is one method to approximate Bayesian inference with a lower computation cost by using dropout at test time [15]. Other methods such as temperature scaling [16] or histogram binning [17] calibrate fully trained network outputs without changing inference. Parameters are learned on the validation set to map network outputs to a true confidence distribution. These methods have the advantage of maintaining inference time and increasing the interpretability of the results without introducing a loss in the accuracy of the model.

### D. Contributions

After an analysis of the data, we determined that although the spectral information and ECG data are useful, the classification can be performed without them if a suitable B-mode image, ROI coordinate location, baseline position, and mode parameter are given. Although the spectrum provides some useful information (and is used by clinical experts when labelling images), there are many variations in the collection of the spectra that make it difficult to use in a network. For example, as shown in Fig. 3, the spectral data can have jumps in the baseline as the user changes the parameters during acquisition. The spectral data is also of variable length which effectively shrinks or expands the features in the output spectral image. Dealing with this would require an even larger dataset since CNNs are not magnitude invariant. To avoid adding unnecessary complexity, we developed a method that does not rely on spectrum data and ECG signal and focused on the integration of the latter four parameters in our classification network. The principle contributions of this work are three-fold:

(1) We show how heatmaps can be used to encode spatial features at the input of CNNs when multi-modal data includes coordinate locations as features. In our work we encode the ROI as a heatmap and append it to the raw B-mode image.

(2) We borrow techniques from multi-task learning to develop a multi-head learning strategy that integrates mode information to prevent misclassifications and reduce network size.

(3) We demonstrate how neural network layers besides just the final layer can be used to define a confidence metric that will disregard many images that differ from the training set. Our method requires no extra trained parameters, uses a fully nonlinear mapping between the output values and the network confidence estimate, and can be dynamically modified at inference time depending on the desired tradeoff between ignored and error rates.

To the best of our knowledge, this is the first work to use CNNs to classify Doppler spectra. We achieve high accuracy on the task, while maintaining a small memory footprint and close to real-time performance. Moreover, several of the methods developed in this work may be applicable to other classification problems, especially in medical imaging.

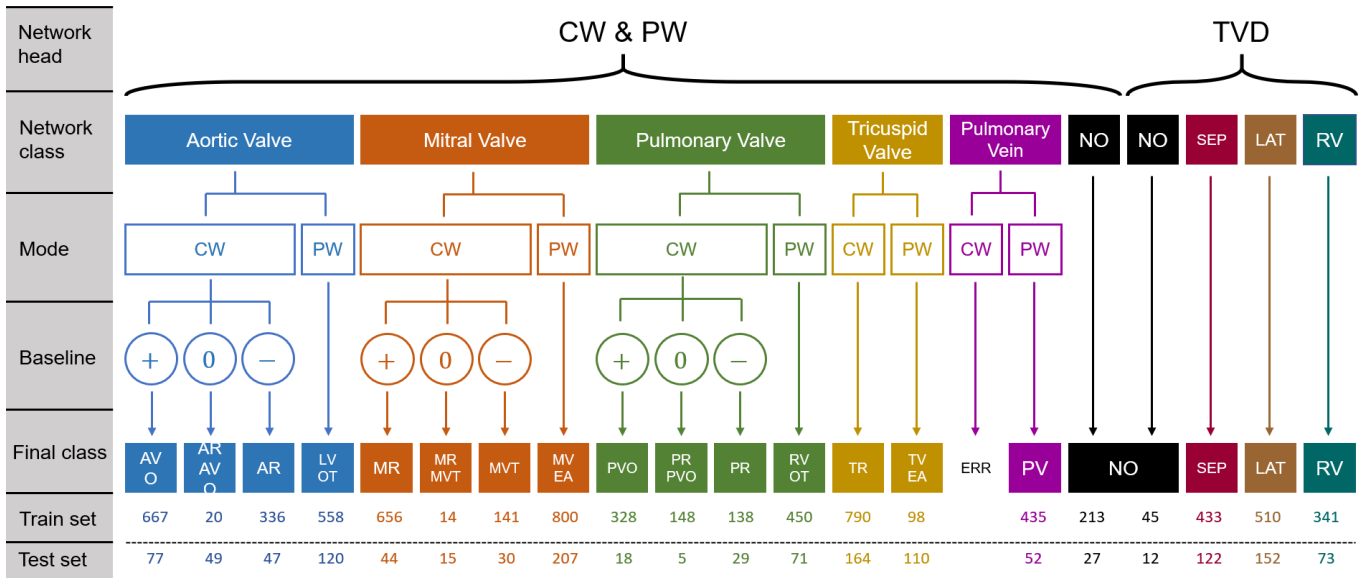

Fig. 5. Class mapping for the network from the 10 classes used by the network to the 19 output classes. The network had two heads, the CW&PW head and the TVD head. Outputs from each head were mapped to final classes using the mode and baseline parameters. Note that there is a separate No Organ (NO) class for both heads, but these are merged to a single class in the output. For the baseline: (-) indicates a baseline in the range [0, 0.5), (0) indicates a baseline at 0.5 and (+) indicates the range (0.5, 1]. If nothing is indicated for mode or baseline, then those parameters are not used for that mapping (all values map to the same class). For example. every image in the TVD head is guaranteed to be mode TVD so the mode is not used in this head. Training and test set sizes are shown below each class. See Fig. 2 for locations of each class and acronym definition. ARAVO, MRMVT, and PRPVO are combinations of AR/AVO, MR/MVT, and PR/PVO respectively.

## II. METHODS

The proposed method performs a classification of the most common Doppler spectrum classes, as described in Sec. A. The proposed method uses only the B-mode image data, the relative ROI position of the Doppler cursor, the baseline direction, and the imaging mode, as described in Sec. B. A multi-head network output approach is used to divide the classification according to the imaging mode, as described in Sec. C. Finally, a confidence metric is defined to avoid misclassifications in the case of low-confidence cases such as for spectra that fall outside of our class set or images with poor quality, as defined in Sec. D.

### A. Network classes

Through conversations with clinical experts we identified 18 of the most common spectrum classes for adults. Three additional types of Doppler spectra do exist but are infrequent in clinical practice and are thus excluded from the current network design. To account for classes not covered in our label set we take two steps. First, to avoid making a classification on images scanned without a visible B-mode image on screen, a no organ (NO) class was added which consisted of images where air and varying amounts of ultrasound gel was scanned. The ROI, baseline, and other parameters were chosen to cover a variety of possible inputs for the NO images. Second, we design a confidence metric in section II D to discard images from other classes. A full discussion of each of the classes is outside the scope of this work, but Fig. 2 shows a diagram of the relative ROI position of each class as well as the acronyms for each class. An outline of each measurement's use and acquisition is available in [1], and reports specific to CW and PW [18], and TVD [19] mode measurements are also available.

### B. Network Input

As shown in Fig. 1, a single Doppler recording is composed of many multi-modal features. An expert observer is easily able to integrate the relevant information and classify the type of spectrum shown. However, it would be unrealistic to expect a network to be able to perform a classification given only an image such as Fig. 1 because some of the most important pieces of information are not emphasized in the image. For example, the region of interest (ROI) is very important to the classification because it indicates the location of the Doppler spectrum within the heart, but it is only a small marker on the image. To mitigate this, we extract all the relevant data individually from each recording. The raw B-mode data is encoded as an image, the ROI as a coordinate location, the relative baseline as a float in the range from 0 to 1 where the default (unchanged) location is 0.5, and the mode is one of CW, PW, or TVD. The non-scanconverted B-mode is extracted as a 512x256 image since the depth dimension is usually much larger in the raw data. Note that the non-scanconverted (beam space) data is used directly rather than the scan-converted (probe space) data that is shown to the user, since the added step in the pipeline to scan-convert the images yields no gain in this application where the ROI position relative to the heart structures is the key piece of information. The position of the ROI is extracted relative to the original B-mode image as a coordinate pair, but Liu *et al.* showed CNN's are typically poor at learning mapping between coordinates in cartesian coordinate space and pixel space [20]. However, in landmark detection problems, networks are able to generate heatmaps of likely locations with high accuracy [21]. Intuitively this makes sense since there is a one-to-one mapping from the coordinate space of the input image to the output heatmap. It follows that networks would also converge faster if landmarks at the input are also encoded as heatmaps, so to encode our ROI location we generate a 2D normal gaussian with a standard deviation of 10 pixels centered at the ROI coordinate. The heatmap is generated in 512x256 resolution to match the original raw data and then appended to the input image as an additional channel before rescaling both to 256x256. This has the effect of squeezing the gaussian vertically, which allows the expected spatial distribution of the landmark to more closely match the physical dimensions of the raw data. An example heatmap is shown in Fig. 4.

## C. Network Output

### 1) Multi-head Network

The mode parameter can be used to linearly split our set of classes into unique sets of non-overlapping classes (except for the no-organ synthetic class). Therefore, one straight-forward approach is to train a different network for each mode. However, this approach doubles the memory footprint of any implementation, which is a downside for integration into a resource-constrained environment. An alternative solution is to frame this as a multi-task learning problem. Multi-task learning integrates the information from several related tasks into a single network by implementing a separate softmax classifier (or classification "head") for each task. Often in multi-head networks, information from one task improves performance on another and the approach has proven to be successful in a variety of deep learning applications [22]. Our method is slightly different from a multi-task approach since the task is the same for each head and each input example will only result in feedback from a single head, so we instead term this a multi-head network. The architecture of our multi-head network is shown in Fig. 4.

With this design choice we exploit the information about the different modes by including one head and loss function for each mode. Input to the classification head for each mode are the values from the last fully connected layer for the classes that belong to that mode. During training, we backpropagate loss only from the head with a mode matching that sample. During inference, only the values from the relevant mode are read at the output.

### 2) Mapping Scheme

As shown in Fig. 2, the B-mode image and ROI cursor location can overlap for several of the classes. In these cases,

the only method for determining which class is present is the mode and relative baseline position extracted from the spectrum. Since these parameters linearly separate the classes and are parameters that can be read from the file we chose to introduce a post-processing mapping scheme rather than feeding them into the network. This scheme enables us to use classes that are based solely on the ROI position and B-mode image. Because of position overlap, the CW and PW classes are merged into a single head and the TVD classes are a separate head. The original 18 classes are mapped to 8 new classes as shown in Fig. 5. The multi-head approach requires a separate NO class for each head since such an image can occur in either mode, so we finish with 10 classes output from the network. One possible error is introduced in this scheme when a CW image is classified as a Pulmonary Vein (PV). In preliminary experiments this was never an issue, but occurrence in a clinical setting would require manual re-classification. Merging the classes with this mapping scheme also increased the training and test set sizes for each trained network class. This is an important consideration since several of the original smaller classes did not have enough images for a network to properly converge.

## D. Confidence metric

Correct classifications from the network will yield significant time savings for clinical users by automatically launching the measurement tool associated with that Doppler class, where available. However, incorrect classification comes with a cost as the user will then have to navigate back in the menu before selecting the correct measurement. As automation continues to permeate clinical workflows this cost may become larger as an initial misclassification could trigger unrelated measurements and automated tools. Moreover, there may frequently be images in a clinical setting that are far different from those seen during training. Thus, it is important for the network to have a bail-out mechanism on images with high uncertainty. In our approach, we use the last fully connected layer before the softmax classifier: the "pre-softmax" layer. This layer was chosen because we could easily extract raw network estimates for all classes before they were distorted by the multiple heads. The output of the pre-softmax layer for each example in the training set is recorded after the network weights are trained and frozen. The training set is used because the validation set is not large enough to provide statistical significance. The recorded values are divided into quantiles. That is, rather than learning a mapping from outputs to true confidence (as was done in [16] and [17]), we find a series of cutoff values for each confidence level. During test time, the quantile is set based on the desired tradeoff between error rate and ignored rate. The maximum output value is found as usual, but if the pre-softmax value for that class falls below the given threshold then the image is labeled as low confidence and ignored.

| # | Network | Information | Output | Accuracy | Estimated Size (MB) | Inference time (ms) |
|---|---------|-------------|--------|----------|---------------------|---------------------|
| E1 | ResNet18 | Image | Separate | 67.1 % | 210 | 4.32 |
| E2 | ResNet18 | Image+Heatmap | Separate | **96.8 %** | 210 | 4.32 |
| E3 | ResNet18 | Image+Heatmap | One head | 91.1 % | 105 | 4.18 |
| E4 | ResNet18 | Image+Heatmap | Train: one head Test: two heads | 96.2 % | 105 | 4.29 |
| E5 | ResNet18 | Image+Heatmap | Two heads | 96.2 % | 105 | 4.32 |
| E6 | ResNet34 | Image+Heatmap | Two heads | 96.2 % | 178 | 7.45 |
| E7 | ResNet50 | Image+Heatmap | Two heads | 95.8 % | 376 | 10.8 |
| E8 | SqueezeNet | Image+Heatmap | Two heads | 93.4 % | 54 | 4.06 |

Table 1. Results of network experiments. In the Output column "separate" refers to training two separate networks, "one head" refers to a single network with all classes, and "two heads" refers to the multi-head approach detailed in section II C. Memory size estimates come from the pytorch-summary package *[26]*.

## III. RESULTS

### A. Dataset

The training and validation datasets consisted of exams previously collected by GE Healthcare for internal tool development. All exams were fully anonymized and came from a variety of clinical sites. Exams were collected to try to maintain a high number in each class, but more images were naturally available for classes that occur more frequently in clinical practice (e.g. MR) than those that occur infrequently (e.g. MRMVT). Since these classes were joined together by our mapping scheme before network training, class imbalance was not an issue for the network. The final set was 3362 images where individual class sizes are shown in Fig. 5. The images were split 90%/10% for each class into training/validation sets respectively.

The test set was separately collected from two institutions that did not have any exams in the training dataset. This was done for two reasons. First, since images are fully anonymized it is impossible to guarantee that two images from the same institution are not from the same patient. It is crucial for accurate test statistics that the training and test sets contain unique patients. Second, every institution has slightly different acquisition practices and patient populations leading to small differences in the distribution of the images. Thus, to get a result that reflects real performance "in-the-wild" it is important to test on data from a separate institution. The test set contained 1424 images and class distributions are also shown in Fig. 5. All images were labeled by a clinical expert experienced in Doppler spectrum analysis and reviewed for accuracy by two other experts.

While gathering the training and validation sets, there were 298 images that had insufficient image quality for an expert to classify them. These images were set aside as the "unknown" set to analyze the confidence metric. Additionally, 30 images were identified that belonged to spectrum classes not included in this network because they appear infrequently in clinical practice. These images were also put aside as the "extra" set to analyze the confidence metric.

### B. Implementation details

All pre-processing, training, and testing were carried out on Ubuntu machines, each with Python 3.6, PyTorch 0.4, and an NVIDIA Titan X GPU. The validation set was used to evaluate preliminary network architectures. All networks were retrained using both the training and validation sets before they were evaluated with the final test set. The images were normalized to [0,1] and mean-centered based on the mean values of the training set, subtracting 0.3 for the B-mode image and 0.0068 for the heatmap image. For augmentation we used random image crops of 224x224 pixels during training. During validation and testing we used center-cropping.

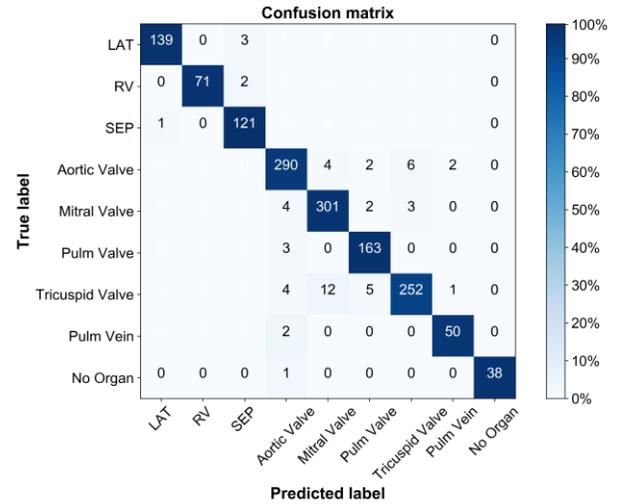

Fig. 6. Confusion matrix on the final test set using the ResNet18 multi-head approach. The boxes where no number is shown will never be misclassified because separate classification heads are used. Colors are normalized to class size (percentages).

### C. Experiments

To evaluate the effects of our design decisions we constructed a series of experiments with results shown in Table 1. First, we evaluated the effect of adding the ROI heatmap. To do this we trained using only the B-mode image as an input (E1 in Table 1) as a baseline and then trained with the ROI heatmap appended to the B-mode image (E2 in Table 1). In both cases separate networks were trained for each mode (CW&PW vs. TVD). Both experiments used a ResNet18 network (architecture shown in Fig. 4) [23]. The ResNet architecture was chosen because it has achieved high accuracy on a variety of classification tasks. Specifically, ResNet18 has a smaller footprint than other networks and less parameters which helps avoid overfitting on data-limited tasks. As expected, results showed a significant improvement when the heatmap was included, with overall classification accuracy increasing from 67.1% for E1 to 96.8% in E2.

Second, we tested the effect of the multi-head approach. As a baseline approach we trained one network with a single classification head on all 9 classes (E3 in Table 1). There are 9 classes instead of 10 here because with a single classification head only one NO class is needed. Results showed a significant drop in accuracy compared to E2 indicating that the network is not able to fully detect the mode by itself. However, the memory footprint was also cut in half. Our goal was to use the multi-head approach to obtain a network with the same accuracy as two separate networks (E2) but the memory footprint of a single network (E3). To test this, we trained a network with the 9 classes and a single classification head but used the multi-head approach during test time (E4 in Table 1). Next, we tried a single network using the multi-head approach during both training and testing (E5). Experiments E2, E3, E4, and E5 all use the same input information, but with different methods of integrating the mode information.

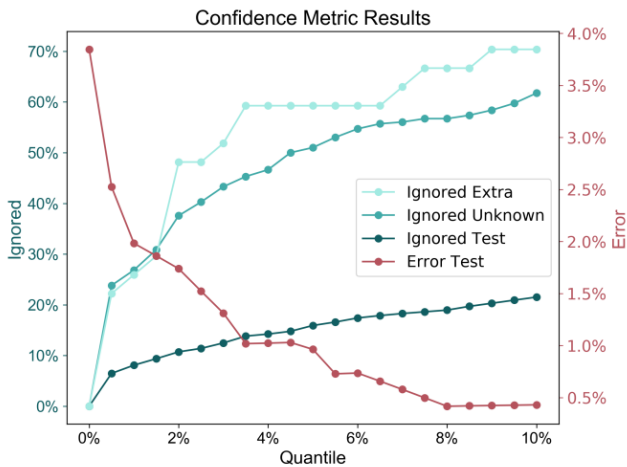

Figure 7 Results of confidence metric experiments with ignored rates (green lines) on the left axis and error rate (red lines) on the right axis. Ignored refers to the percentage of images which the network did not label. The "unknown" images are those which were unidentified during labeling. The "extra" images came from classes not included in this network. For the "unknown" and "extra" sets an ideal confidence metric should ignore all images. For the test set an ideal confidence metric should ignore as few images as possible while reducing error rate (every misclassified image should be ignored).

Results showed equivalent results (96.2%) for E4 and E5, with a small decrease in accuracy from E2 to E4 and E5. This was somewhat surprising since E5 consistently obtained the best results on the validation set but indicates that it may not generalize as well. The multi-head approach was still able to achieve almost equivalent results with half of the memory footprint. Since our network needs to operate in a resource-constrained environment and results were similar, we chose to use the multi-head approach for the next set of experiments.

Third, we tested how results would change with the chosen network size. To test if deeper networks would increase accuracy we used larger versions of ResNet: ResNet34 (E6) and ResNet50 (E7). To test if similar accuracy could be maintained with a smaller network we used SqueezeNet (E8). SqueezeNet is a network designed for high performance on classification tasks in resource-constrained environments [24]. Results showed that ResNet34 maintained similar results, while there was a small decrease when moving to ResNet50. Additionally, the memory footprint and inference time significantly increased with these architectures. This indicates that deeper networks may overfit to the smaller dataset used here and come with increased memory size and inference time. There was a larger performance degradation when using SqueezeNet (93.4%).

Overall the optimal approach for implementation considering accuracy and network size was the ResNet18 multi-head approach (E4 and E5). Results were almost identical between E4 and E5, but E5 has a simpler implementation so was chosen for final testing. The confusion matrix on the test set for E5 is shown in Fig. 6 Tricuspid Valve was the lowest accuracy class with 91% (all other classes were over 95%).

## D. Confidence Metric

To test the validity of the proposed confidence metric we extracted the pre-softmax set of cutoff values on the training set and extracted quantile cutoff limits for each class from 0%-10% in 0.5% step sizes. The quantile is the ignored percentage on the training set: a quantile of 5% indicates that the 5% of images with the lowest pre-softmax values would be ignored while a quantile of 0% would never ignore any images and correspond to a network with no confidence metric.

The quantiles were used when running inference on the test set of images as well as the "unknown" and "extra" sets that were put aside during labeling. For each set, we swept the quantile value and recorded the ignored rate. For the test set the error rate was also recorded. Results are shown in Fig. 7 with ignored rates on the left axis and error rate on the right.

The confidence metric results likely indicate some overfitting, but also may indicate that the training and test sets came from different distributions. If the network is not overfit and the images are from the same distribution, the quantile should map 1:1 to the ignored test percentage but results show at the 0.5% quantile more than 6% of the test images are ignored. Some difference in distribution is expected given that the images came from separate clinics. As hoped, the "unknown" and "extra" image sets are ignored at a much higher rate: >20% of images from both sets are ignored at the 0.5% quantile. The ignored rate shows an approximately logarithmic relationship with increasing quantile values. At the 10% quantile 21% of the test set images, 70% of the "extra" images and 61% of the unknown images are ignored. The error rate drops with increasing quantile levels, reaching the 1% mark at the 3.5% quantile.

To validate the use of the pre-softmax layer, quantile limits were also set with several other methods. We implemented a MC-dropout [15] version of our model following the approach in [25] where 50% of the neurons from the last fully connected layer were dropped during each inference run. Each MC-dropout model was run 100 times and we set quantile limits for values from the pre-softmax layer and softmax layer of the normal model, and from the mean and variance of the pre-softmax and softmax layers from the MC-dropout model. Results were similar for all implementations, with a slightly higher ignored rate for all sets when using the pre-softmax layer from either model. These results indicate that the choice of how to extract quantile values does not play a large role in the resulting confidence metric.

## IV. Discussion

Our results indicate that highly accurate Doppler spectrum classification is possible in echocardiography *without* using the spectrum data in the training. In this case the result follows from the fact that each class (after using the mapping scheme) can be assigned to a unique physical location within the heart. However, since the spectra can be acquired from a variety of different views this position can vary significantly within the image. Accurate classification requires understanding of both the B-mode image and the ROI location. While highly accurate results have already been achieved on view recognition tasks in echocardiography (e.g. [7], [8]),) our results indicate heatmap encoding is an effective way to pass location information into CNNs. With this structure CNNs can achieve high accuracy on this task.

Since deep-learning algorithms deployed in clinical settings must frequently compete for resources, we also analyzed how we could decrease resource utilization. We demonstrated a multi-head classification could be used to reduce the memory footprint when the classification task can be split into separate problems. The network was able to maintain similar accuracy levels to those achieved by separate networks and much higher accuracy than a single network trained with all classes. Our implementation achieved sub 5ms inference time, indicating near real-time performance.

Misclassifications can be costly in a medical setting. It can lead to confusion when analyzing patient data and mistrust in artificial intelligence-based tools. To attempt to reduce misclassifications, we took several measures. First, we included a No Organ (NO) class in the training dataset to avoid classifying images of air and gel into another class. Second, we analyzed the training set output values from the last fully connected ("pre-softmax") layer for each class and set cutoff limits. Images with a score below the cutoff were discarded rather than classified. Overall, results indicated the proposed confidence metric and variations of it can significantly reduce the error rate of the network. The confidence metric also ignores the "unknown" and "extra" image sets at a much higher rate than those that are part of the test set. Moreover, this implementation allows a user to easily set the quantile limit depending on the desired tradeoff between the error percentage and ignored percentage.

## V. Conclusion

In this work we demonstrated a CNN-based method for the automated classification of Doppler spectra including a confidence metric to discard images with high uncertainty. We showed notable performance gain on the task by encoding the ROI as a heatmap and appending it as a channel to the input and introducing a post-processing mapping scheme to simplify the problem. Future work will focus on extending the training dataset to continue to improve accuracy. Another possible method for increasing accuracy is to use multi-task learning with cardiac view recognition since the tasks are similar. We will also explore other possible methods of encoding network confidence without sacrificing performance.